\documentclass[aps,prb,showpacs,superscriptaddress]{revtex4}

\usepackage[cp1251]{inputenc}
\usepackage[russian,english]{babel}
\usepackage{graphicx}
\usepackage{color}
\usepackage{bm}
\usepackage{amssymb}
\usepackage{amsmath}

\newcommand{\be}{\begin{equation}}
\newcommand{\ee}{\end{equation}}
\newcommand{\ben}{\begin{equation*}}
\newcommand{\een}{\end{equation*}}

\begin{document}
\title{To the problem of electron-hole  bound state in transition-metal dichalcogenides}


\begin{abstract}
The interacting electron and hole in transition-metal dichalcogenides is considered. For investigation of the interaction between electron and hole we obtain the Bethe-Salpeter equation for two interacting Dirac particles. The dependence of a few lowest binding energies of electron and hole on the interaction constant for different potentials is found. We demonstrate that the behavior of the potential at small distances significantly affects on the values of the binding energies. For small interaction constant we have developed the perturbative method of the binding energy calculation. For the largre interaction constant the binding energies are found numerically. The critical values of the interaction constant for the Coulomb potential and exponential potential are found.
\end{abstract}

\hskip 2cm
\date{\today}

\author{P.~A.~Krachkov}
\email[E-mail: ]{}
\affiliation{Budker Institute of  Nuclear Physics of Siberian Branch Russian Academy of Sciences, Novosibirsk, 630090 Russia}
\author{I.~S.~Terekhov}
\email[E-mail: ]{I.S.Terekhov@gmail.com}
\affiliation{Budker Institute of  Nuclear Physics of Siberian Branch Russian Academy of Sciences, Novosibirsk, 630090 Russia}


\maketitle

\section{ Introduction}

In 2004 the famous article devoted to the study of the first two-dimensional material  appeared \cite{Novoselov:2004}. This article opens a new direction of experimental and theoretical investigations of two-dimensional materials. After creation of the graphene many different two-dimensional and quasi two-dimensional materials were investigated both experimentally and theoretically. The interesting family of the materials is the transition-metal dichalcogenides. Some of the materials are the semiconductors which have the value of the band gap of the order of $2~eV$, see eg. \cite{Zhu:2011}. The investigation of the energies of the exciton states in such materials is  an interesting problem, from both experimental and theoretical point of view. The experimental investigation of the exciton spectrum and comparison of the spectrum with the one predicted by the theory gives the understanding of the nature of the electron-electron interaction in the considered materials. It allows us to predict the properties of the similar materials. From theoretical point of view the problem of electron-hole interaction is interesting due to the non-triviality of including the interaction to the model. There were many attempts to include the interaction to the system, see e.g.  \cite{Rodin:2013,Zhou:2015,Trushin:2016,Trushin:2018,Trushin:2019}. In the Ref. \cite{Trushin:2016} to obtain the Hamiltonian of the system of interacting electron and hole the authors perform the transformation  of the Hamiltonian for  electron-hole system without interaction to the block diagonalized form, Then the authors expand  obtained Hamiltonian in the assumption that the kinetic energy is much less than the band gap. After that the Coulomb interaction was added to the expanded Hamiltonian. In Ref. \cite{Trushin:2018} the Hamiltonian of the system was chosen as the sum of three terms. First term
corresponds to  the  free particle Hamiltonian with  reduced mass of electron and hole. Second term corresponds to the interaction potential, and the third therm corresponds to the corrections related to the Berry curvature. In Ref. \cite{Rodin:2013} to consider the exciton spectrum and to investigate the critical behavior of the system the two-dimensional modified Dirac equation  was considered. In the modified equation the momentum operator is doubled and the Coulomb field is added. In Ref. \cite{Zhou:2015} to describe the bound state of electron and hole the three types of equations were considered. First equation contains sum of free particle Hamiltonian with  reduced mass and interaction potential. The second term is related to the Berry curvature. Second type of the equation corresponds to the expansion of the Hamiltonian which is sum of two  Hamiltonian for free Dirac particles and interaction potential  in the case when the binding energy is much less than band gap. The third equation is some modification of the Bethe-Salpeter equation (BSE).

Different approaches to the problem demonstrate the difficulty of the  including the interaction between electrons  to the system. The difficulty is related with the appearance of the electron-hole excitations in the intermediate states \cite{Berestetski,Itzykson}. The problem was solved in the framework of the quantum field theory. The existence of the bound states of the electron and hole manifests itself in the form of singularities in the two particle scattering amplitude. The equation which describes the singularities of the amplitude is the BSE. Therefore, to find the exciton spectrum it is necessary to obtain the BSE for the system of interacting electron and hole. The BSE for the semiconductors was obtained in Ref. \cite{Glinskii:1987}. In Ref. \cite{Scharf:2019} the authors solved numerically the BSE and found the exciton spectrum. To obtain the spectrum the authors took into account the random phase approximation for the potential and the corrections related with the self-energy operator. However, in Ref. \cite{Scharf:2019} the one particle Green's function corresponds to a particle with parabolic dependence of the energy on the momentum. In transition-metal dichalcogenides the one particle excitations are described by the two-dimensional Dirac equation. Therefore, the wave function should have two component and the dependence of the energy on the momentum is not parabolic. The approximation performed in Ref. \cite{Scharf:2019} means the smallness of the kinetic energy of the particle in comparison with band gap. For excitons the kinetic and interaction energies are of the order of band gap, therefore such approximation is not correct. Thus, further study of the BSE for the interacting electron and hole in the transition-metal dichalcogenides is necessary.

In the present paper we obtain the BSE for the system of intravalley interacting electron and hole in the transition-metal dichalcogenides in the leading order in the interaction potential. We investigate the solutions of the obtained equation for different values of the interaction constant and for different types of the potentials. For the Coulomb potential we find the analogue of the non-relativistic approximation for the equation which can be used in the case of small interaction constant. We demonstrate that this equation differs from that obtained in Ref.\cite{Zhou:2015}. We also find numerically the critical value of the interaction constant, i.e. the value of the interaction constant at which the ground state energy of the interacting electron and hole reach the valence band. To investigate the behavior of the solutions in the vicinity of the critical values of the interaction constant, we consider the localized potential which does not contain any singularities.

The paper is organized as follows. In the Section \ref{SecModel} we describe the model, obtain the BSE for the electron and hole. In the Section \ref{SecMallIntConst} we consider the approximation of the BSE at small interaction constant. We demonstrate that standard method of the expansion of the BSE for the Coulomb potential leads to the appearance of the non-integrable corrections to the interaction potential \ref{SingularPotential}. In the Subsection \ref{CorrectExpansion} the correct expansion of the BSE is obtained. In the Section \ref{BetheSalpeterNumeric} the exact numerical solution for the BSE is obtained. In this section we compare the exact solution and the one obtained using the perturbation theory. We also consider the critical values of the potential. In the Conclusion we discuss the obtained results.

\section{Model\label{SecModel}}
The starting point of our consideration is the one particle Hamiltonian suggested in \cite{Xiao:2012}:
\begin{eqnarray}\label{Hamiltonian}
\hat{H}_\lambda=v_F\left(\tau p_x\sigma^x+p_y\sigma^y\right)+\frac{\Delta}{2}\sigma^z-\lambda\tau\frac{\sigma^z-1}{2}s_z,
\end{eqnarray}
where $(\sigma^x,\sigma^y,\sigma^z)$ are the Pauli matrices, $\bm p =(p_x,p_y)$ is the momentum operator, $s_z$ is the Pauli  matrix for spin, $\tau$ is the valley index, the parameter $\lambda$ denote the spin-orbit coupling parameter. In Eq. \eqref{Hamiltonian} we introduced the Fermi velocity parameter $v_F=a t/\hbar$, $a$ is the lattice constant, $t$ is the effective hopping integral, $\hbar$ is the Planck constant. The values of the parameters $\Delta$, $\lambda$, $t$, and $a$ can be found in the Table 1 of  Ref.~\cite{Xiao:2012}. Below we set $\hbar=v_F=1$.

Let us consider the eigenvalues $\epsilon_{\tau,s_z}$ of the Hamiltonian \eqref{Hamiltonian} for different parameters $\tau$ and $s_z$. We have
\begin{eqnarray}
&& \epsilon^{(\pm)}_{1,1} (p)=\epsilon^{(\pm)}_{-1,-1} =\frac{\lambda}{2}
\pm
\sqrt{p^2+\frac{(\Delta-\lambda)^2}{4}},\\
&& \epsilon^{(\pm)}_{1,-1}(p)=\epsilon^{(\pm)}_{-1,1} =-\frac{\lambda}{2}
\pm
\sqrt{p^2+\frac{(\Delta+\lambda)^2}{4}}.
\end{eqnarray}
So, for the valley corresponding to $\tau=1$ the gap between conduction and valence band is equal to $\Delta-\lambda$. The conductance band has lower  bound which corresponds to the $\epsilon^{(+)}_{1,-1}(p)$. The valence band has upper bound  which corresponds the $\epsilon^{(-)}_{1,1}(p)$. For the valley corresponding to the $\tau=-1$ the gap between conduction and valence band is also equal to $\Delta-\lambda$. For this case, we have lower bound  $\epsilon^{(+)}_{-1,1}(p)$ for conductance band, and upper bound $\epsilon^{(-)}_{-1,-1}(p)$ for the valence band. Since we have eight branches in the energy spectrum, after second quantization procedure we obtain eight different particles which correspond to two valleys, two spins, and two branches for each valley and spin.  This particles have different masses $m_{1,2}=\frac{\Delta\pm\lambda}{2}$. The branches $\epsilon^{(+)}_{\tau,s_z}$ correspond to the quasiparticles (electrons), whereas the branches $\epsilon^{(-)}_{\tau,s_z}$ correspond to the anti-particles (holes). One can check that the ratio $\lambda/\Delta\ll 1$, therefore below in the present paper we will consider the case $\lambda=0$. In this case the masses $m$ of the quasiparticles equal to $\Delta/2$.

To find the energy spectrum of excitons (bound state of an electron and hole) it is necessary to introduce the interaction in the Hamiltonian \eqref{Hamiltonian}. 

In Refs.\cite{Rodin:2013,Zhou:2015,Trushin:2016,Trushin:2018,Trushin:2019} the starting point of consideration is the expansion of the Hamiltonian in the vicinity of the electron-hole pair energy which equals to $2m$.  Then the authors add the interaction potential $V(r)$, and some terms related with Berry curvature, see, e.g., Refs. \cite{Zhou:2015,Trushin:2018}. Such consideration is similar to the non-relativistic expansion in quantum electrodynamics. It is well known that this expansion is valid when the ratio $(2m-E_{exc})/m\ll 1$, where $E_{exc}$ is the electron-hole pair energy. However, as can be seen from experimental data \cite{Chernikov:2014,Kumar:2014}, the  ratio $(2m-E_{exc})/m$ is of the order of unity for the ground state for such materials as $WS_2$ and $WSe_2$. It means that the effective interaction constant between quasiparticles is of the order of unity. For such interaction parameter the analogue of the non-relativistic expansion is not applicable because the electron-hole excitations in the intermediate state can significantly change the interaction, see \cite{Berestetski,Itzykson}. Therefore applicability of this expansion is disputable.

In quantum field theory the method of the energy calculation for the bound states is developed. It based on the investigation of the singularities in the two-particle scattering amplitude.  The BSE \cite{Itzykson}  describes singularities in the two particle scattering amplitude which related to the bound states of electron and hole (electron and positron in quantum electrodynamics), therefore we should obtain the equation for the electron-hole system in the transition metal dichalcogenides.

For simplicity, below we consider the case $\tau=1$ and neglect spin-orbit interaction. In this case the Hamiltonian \eqref{Hamiltonian} has the form:
\begin{eqnarray}\label{Hamiltonian1}
\hat{H}=\bm \sigma\cdot\bm p+m\sigma^z,
\end{eqnarray}
where $\bm \sigma=(\sigma_x,\sigma_y)$, $\bm p=(p_x,p_y)$. In the Hamiltonian \eqref{Hamiltonian1} we omit spin index, since it diagonal in the spin variable.  Using the technique described in Ref.\cite{Itzykson} we derive the BSE for interacting electron and hole in the leading order in the interaction potential:
\begin{eqnarray}\label{BeSalint}
\Psi^{i,j}(\varepsilon_1,\bm p_1|\varepsilon_2,\bm p_2)=-iG_e^{i,n}(\varepsilon_1,\bm p_1)G_h^{j,l}(\varepsilon_2,\bm p_2)
 \int \dfrac{d\bm q\, d\omega}{(2\pi)^3}{ V(\bm q)}\Psi^{n,l}(\varepsilon_1+\omega,\bm p_1+\bm q|\varepsilon_2-\omega,\bm p_2-\bm q),
\end{eqnarray}
where $V(\bm q)$ is the Fourier transform of the electron-electron interaction potential $V(r)$, $G_{a}^{i,m}(\varepsilon,\bm p)$ is the one-particle Green's function:
\begin{eqnarray}
G_a^{i,j}(\varepsilon,\bm p)=\frac{\varepsilon+\bm{\sigma}_{a}^{i,j}\cdot\bm{p}+m(\sigma_a^z)^{i,j}}{\varepsilon^2-\bm{p}^2-m^2+i0}\,,
\end{eqnarray}
where index $a$ enumerates electron and hole. The imaginary part of the Green's function corresponds the Fermi energy equals to zero (center of the band gap). The function $\Psi$ depends on the two energies $\epsilon_1$ and $\epsilon_2$, but the wave function of an electron-hole pair should depend only on one energy parameter.  So the quantity $\Psi$  can not be treated as the wave function. To obtain the equation for the wave function we use standard prescription\cite{Itzykson}. We change the variables $\varepsilon_1= E/2+\Omega$, $\varepsilon_2= E/2-\Omega$ and perform the integration over $\Omega$ in the both sides of Eq.~\eqref{BeSalint}. As result we obtain:
\begin{align}\label{BeSalint1}
&\tilde{\psi}^{i,j}(E,\bm p_1,\bm p_2)=-i\int \dfrac{d\Omega}{2\pi} G_e^{i,n}\left(\frac{1}{2}E+\Omega,\bm p_1\right)G_h^{i,l}\left(\frac{1}{2}E-\Omega,\bm p_2\right)\nonumber\\
&\times \int \dfrac{d\bm q}{(2\pi)^2} V(q)\,\tilde{\psi}^{n,l}(E,\bm p_1+\bm q,\bm p_2-\bm q)\,.
\end{align}
Here we introduce the wave function of the electron-hole pair:
$$\tilde{\psi}(E,\bm p_1,\bm p_2)=\int \frac{d\Omega}{2\pi}\Psi\left(\dfrac{1}{2}E+\Omega,\bm p_1|\dfrac{1}{2}E-\Omega,\bm p_2\right).$$
Below we consider the case $\bm p_1=-\bm p_2=\bm p$. Performing the integration over $\Omega$  in  Eq.~\eqref{BeSalint1},  we finally  obtain the equation for the wave function $\tilde{\psi}(E,\bm p,-\bm p)=\psi(E,\bm p)$:
\begin{eqnarray}\label{BetheSalpEq}
&&(E-(\bm{\sigma_e}-\bm{\sigma_h})\cdot\bm p-(\sigma_h^z+\sigma_e^z)m)\psi(E,\bm p)=\left(\Lambda^{--}(\bm p)-\Lambda^{++}(\bm p)\right)
\int \dfrac{d\bm q}{(2\pi)^2} V(|\bm p-\bm q|)\psi(E,\bm q)\,,
\end{eqnarray}
where we omit the indexes for $\sigma$ matrices,  the operators $\Lambda^{\pm\pm}$ have the form:
\begin{equation}\label{Lambdapmpm}
\Lambda^{\pm\pm}(\bm p)=\Lambda_e^{\pm}(\bm p)\Lambda_h^{\pm}(-\bm p)\,,
\end{equation}
\begin{equation}\label{Lambdapm}
\Lambda_a^{\pm}(\bm p)=\frac{\omega(\bm p)\pm\left(\bm \sigma_a\cdot\bm p+m\sigma_a^z\right)}{2\omega(p)}\,.
\end{equation}
Here $\omega(\bm p)=\sqrt{p^2+m^2}$. One can check that the equation \eqref{BetheSalpEq} is similar to the BSE for electron-positron pair in quantum electrodynamics \cite{Itzykson}. This is not surprising since the Hamiltonian \eqref{Hamiltonian1} is the two-dimensional analogue of the Dirac Hamiltonian of the electron. In the coordinate space the Eq. \eqref{BetheSalpEq} has the form:
\begin{eqnarray}\label{BetheSalpEqCoord}
&&\left(i\frac{\partial}{\partial t}-(\bm{\sigma_e}-\bm{\sigma_h})\cdot \hat{\bm p}-(\sigma_h^z+\sigma_e^z)m\right)\psi(t,\bm r)=\int d\bm R\left[ Q^{--}(\bm r-\bm R)-Q^{++}(\bm r-\bm R)\right]V(R)\psi(t,\bm R),
\end{eqnarray}
where $\hat{\bm p}=-i\left(\frac{\partial}{\partial x},\frac{\partial}{\partial y}\right)$ is the momentum operator, and  $Q^{\pm\pm}(\bm R)$ is a Fourier transform of  $\Lambda^{\pm\pm}(\bm p)$,
\begin{eqnarray}
Q^{++}(\bm R)-Q^{--}(\bm R)=\left[-i\frac{\partial}{\partial\bm r }(\bm{\sigma_e}-\bm{\sigma_h})+m(\bm{\sigma_e}+\bm{\sigma_h})\right]\frac{1}{4\pi R}e^{-m R}\,.
\end{eqnarray}
The Eq. \eqref{BetheSalpEqCoord} has the conserving density $\rho$:
\begin{eqnarray}
\rho=\int d\bm r \left(|\psi^{++}(\bm r)|^2-|\psi^{--}(\bm r)|^2\right).\label{Norm}
\end{eqnarray}
Where $\psi^{\pm\pm}(\bm r)=\int d\bm R\, Q^{\pm\pm}(\bm r-\bm R)\psi(t,\bm R)$.
The density $\rho$ can be negative; therefore, we can not treat it as the probability density, but we can consider it as the quantity which proportional to the charge density, see Ref. \cite{Itzykson}.

\section{Small interaction constant approximation\label{SecMallIntConst}}
Let us demonstrate that the BSE \eqref{BetheSalpEq} can be transformed to the Schr\"odinger equation in the case of small interaction constant. It is convenient to present the function $\psi(E,\bm p)$ in the following form:
\begin{eqnarray}
\psi(E,\bm p)=f(\bm p)|1,1\rangle + h(\bm p)|1,-1\rangle + g(\bm p)|1,0\rangle + d(\bm p)|0,0\rangle,\label{Psi1}
\end{eqnarray}
where $|\Sigma,\Sigma^z\rangle$ is the eigenvectors of the operators $(\bm \sigma_e+\bm \sigma_h)^2$ and $(\sigma^z_e+\sigma^z_h)$: $(\bm \sigma_e+\bm \sigma_h)^2|\Sigma,\Sigma_z\rangle=(\Sigma^2+\Sigma-(\Sigma^z)^2)|\Sigma,\Sigma^z\rangle$, $(\sigma^z_e+\sigma^z_h)|\Sigma,\Sigma^z\rangle=\Sigma^z|\Sigma,\Sigma^z\rangle.$
We substitute the function $\psi$ in the form \eqref{Psi1} to the Eq. \eqref{BetheSalpEq} and obtain the system of equations:
\begin{eqnarray}
(E-2m)f(\bm p)&=&-\sqrt{2}p^-d(\bm p)-\frac{m}{\omega(\bm p)} (\hat{V}f)(\bm p)+\frac{p^-}{\sqrt{2}\omega(\bm p)}(\hat{V}d)(\bm p)\,,\label{fEq}\\
(E+2m)h(\bm p)&=&\sqrt{2}p^+d(\bm p)+\frac{m}{\omega(\bm p)} (\hat{V}h)(\bm p)-\frac{p^+}{\sqrt{2}\omega(\bm p)}(\hat{V}d)(\bm p)\,,\label{hEq}\\
E d(\bm p)&=&\sqrt{2}p^-h(\bm p)-\sqrt{2}p^+f(\bm p)+\frac{p^+}{\sqrt{2}\omega(\bm p)} (\hat{V}f)(\bm p)-\frac{p^-}{\sqrt{2}\omega(\bm p)}(\hat{V}h)(\bm p)\,,\label{dEq}\\
E g(\bm p)&=&0,
\end{eqnarray}
where $p^{\pm}=|\bm p| e^{\pm i\phi_{\bm p}}$, $\phi_{\bm p}$ is the angle of the vector $\bm p$,
\begin{equation}
(\hat{V}\phi)(\bm p)=\int \frac{d\bm q}{(2\pi)^2} V(\bm p-\bm q)\phi(\bm q).
\end{equation}
One can see, that the component $g(\bm p)$ is equal to zero in the case of the nonzero energy $E$; therefore, below we omit the component $g$ from the consideration. The expansion of the system \eqref{fEq}-\eqref{dEq} at small interaction constant is similar to the non-relativistic expansion of the Dirac equation in quantum electrodynamics \cite{Berestetski}.

For small interaction constant (the smallness of the potential will be considered latter) we imply that the  bound states energies $E$ obey the condition $|E-2m|\ll m$. For such energies we imply that the characteristic values of the momentum and potential obey the conditions \cite{Berestetski}:
\begin{eqnarray}
&&\frac{p^2}{m}\ll m,\label{ConditonForP}\\
&&\left |\frac{\int d\bm p f^+(\bm p)(\hat{V}f)(\bm p)}{\int d\bm p |f(\bm p)|^2}\right |\ll m.\label{ConditonForIntegral}
\end{eqnarray}
In the inequality \eqref{ConditonForIntegral} we also imply that the main contribution to the integrals in Eqs. \eqref{fEq}-\eqref{dEq} comes from the momentum scales $p\ll m$. It means that we perform the calculation under the assumption that the function $f$ decreases fast enough and the contribution of the region  $p\gtrsim m$ to the integral is small. The substitution of the energy $E$ in the form
\begin{eqnarray}
E=2m+\varkappa\label{Ekappa}
\end{eqnarray}
to the Eqs. \eqref{fEq}-\eqref{dEq} gives
\begin{eqnarray}
\varkappa f(\bm p)&\approx&-\sqrt{2}p^-d(\bm p)-(\hat{V}f)(\bm p)\,,\label{fEqFirst}\\
h(\bm p)&\approx&0\,,\label{hEqFirst}\\
d(\bm p)&\approx&-\frac{p^+}{\sqrt{2}m}f(\bm p)\,.\label{dEqFirst}
\end{eqnarray}
Here we retain only the terms of the leading and text-to-leading order in the parameter $|\bm p|/m$. The substitution of the function $d(\bm p)$ in the form \eqref{dEqFirst} to the Eq. \eqref{fEqFirst} gives the
Sch\"odinger equation in the momentum space for the particle with reduced mass:
\begin{eqnarray}\label{SchrodingerK}
\varkappa f(\bm p)&=&\frac{p^2}{m}f(p)-(\hat{V}f)(\bm p)\,.
\end{eqnarray}
Performing the Fourier transform of Eq. \eqref{SchrodingerK} we obtain:
\begin{eqnarray}\label{SchordFirst}
\varkappa f(\bm r)&=&\hat{H}_0 f(\bm r),\\
\hat{H}_0&=&\frac{p^2}{m}-V(\bm r),\label{H0}
\end{eqnarray}
So, we show that at small interaction constant the BSE is similar to the Sch\"odinger equation for the particle which has the mass equals to the reduced mass. The reason of the another sign before the potential in Eqs. \eqref{SchrodingerK}, \eqref{H0} is that the potential $V(\bm r)$ is the electron-electron interaction potential. The condition of weakness of the potential or smallness of the interaction constant means that the lowest binding energy of the electron and hole should be much less than the band gap ($|\varkappa|\ll m$).

Let us consider the solution of the Eq.~\eqref{SchordFirst} in the case of the Coulomb potential: \begin{eqnarray}\label{CoulPot}
V_C(\bm r)=\frac{e^2}{\epsilon r},
\end{eqnarray}
where $e$ is the electron charge, $\epsilon$ is the dielectric constant. For the Coulomb potential the eigenvalues $\varkappa$ of the Eq. \eqref{SchordFirst} have the form:
\begin{eqnarray}\label{KappaFirst}
\varkappa_{n,l}=-\frac{m \alpha^2}{4(n+|l|+1/2)^2},
\end{eqnarray}
where $n$ is the radial quantum number, $l$ is the angular momentum quantum number, $\alpha=e^2/\epsilon$ is the integration constant. The applicability condition $|\varkappa|\ll \Delta\sim m$ means that the parameter $\alpha$ should be much less than unity ($\alpha\ll1$). However,  the experimental results for binding energies demonstrate that the parameter $\alpha$ is of the order of $0.5$ for WS$_2$ ($\Delta=2.41$eV, $\varkappa=0.32$eV), and $0.6$ for WSe$_2$ ($\Delta=2.02$eV, $\varkappa=0.37$eV), see \cite{Chernikov:2014}. It means that for such values of the interaction constant it is necessary to consider the exact equation or, at least, the higher order corrections in the parameter $\alpha$ should be taken into account. The exact solution of the BSE for the ground state and for a few levels with $l=0$ is calculated numerically in the Section~\ref{BetheSalpeterNumeric}. In the present section we find the first correction to the Eq. \eqref{SchordFirst} and to the eigenvalue \eqref{KappaFirst} for $l=0$, and $n=0,1$.

\subsection{Singularities in the Hamiltonian\label{SingularPotential}}
Let us demonstrate that the naive expansion leads to the appearance of the non-integrable operators in the Hamiltonian \eqref{H0}. To find the next-to-leading order correction in $\alpha$ we keep the following terms in the system \eqref{fEq}-\eqref{dEq}, and obtain:
\begin{eqnarray}
\varkappa f(\bm p)&\approx&-\sqrt{2}p^-d(\bm p)-\left(1-\frac{p^2}{2m^2}\right) (\hat{V}f)(\bm p)+\frac{p^-}{\sqrt{2}m}(\hat{V}d)(\bm p)\,,\label{fEqSec}\\
h(\bm p)&\approx&\frac{p^+d(\bm p)}{2\sqrt{2} m}\,,\label{hEqSec}\\
d(\bm p)&\approx&\frac{p^-}{\sqrt{2}m}h(\bm p)-\frac{p^+}{\sqrt{2}m}\left(1-\frac{\varkappa}{2m}\right)f(\bm p)+\frac{p^+}{2^{3/2}m^2} (\hat{V}f)(\bm p).\label{dEqSec}
\end{eqnarray}
The substitution of the the function $h$ in the form \eqref{hEqSec} to Eq. \eqref{dEqSec}, and then the substitution of the result for $d$ to the Eq. \eqref{fEqSec} gives the following equation:
\begin{eqnarray}
\varkappa f(\bm p)=\frac{p^2}{m}f(\bm p)-(V f)(\bm p)-\frac{p^4}{4m^3}f(\bm p)+\frac{p^2}{2m^2}(V f)(\bm p)-\frac{p_-(V p_+ f)(\bm p)}{2m^2}.\label{fFinal}
\end{eqnarray}
Here we keep only the terms which have the necessary accuracy. One can see that the operator in the right-hand side of this equation is not hermitian. The reason for that is the function  $f(\bm p)$ can not be treated as the wave function because the normalization condition \eqref{Norm} for the function $f(\bm p)$ differs from the normalization condition for the wave function, see \cite{Berestetski}. So, to obtain the Hamiltonian we should find the relation between the function $f(\bm p)$ and the Schr\"odinger wave function. The normalization condition \eqref{Norm} in the momentum space in the leading and next-to-leading order in the parameter $\varkappa/m$ has the form:
\begin{eqnarray}
\rho\approx\int d\bm p \left(1+\frac{p^2}{2m^2}\right)|f(\bm p)|^2.
\end{eqnarray}
Therefore, performing the change of the function as
\begin{eqnarray}
f(\bm p)=\left(1-\frac{p^2}{4m^2}\right)\psi_{Sh}(\bm p),\label{fToPsiSh}
\end{eqnarray}
we obtain that the density $\rho$ is expressed through the function $\psi_{Sh}$ as
\begin{eqnarray}
\rho=\int d\bm p |\psi_{Sh}(\bm p)|^2,
\end{eqnarray}
where we keep only the terms which have the necessary accuracy. The last expression coincides with normalization condition for the Schr\"odinger wave function. Finally, we substitute the function $f$ in the form \eqref{fToPsiSh} to the Eq. \eqref{fFinal}, then perform the Fourier transformation and obtain the Shr\"odinger equation in the coordinate space:
\begin{eqnarray}
\varkappa \psi_{Sh}(\bm r)&=&(\hat{H}_0+\hat{H}_1)\psi_{Sh}(\bm r),\\
\hat{H}_1&=&-\frac{p^4}{4m^3v_F^2}\psi_{Sh}(\bm r)-\frac{\hbar^2}{4m^2 v_F^2}(\bm\nabla^2 V)+\frac{i\hbar^2}{2m^2v_F^2}[(\bm\nabla V)\times\bm \nabla]_z.\label{ShrEq}
\end{eqnarray}
Here $\bm\nabla=(\frac{\partial}{\partial x},\frac{\partial}{\partial y})$, $[\bm a\times\bm  b]$ is the vector product.  In the Eq. \eqref{ShrEq} we also recover the dimension. One can see that the operator in the left-hand side of Eq. \eqref{ShrEq} is the hermitian operator, the function $\psi_{Sh}$ obeys the correct normalization condition. Therefore, we can treat the function $\psi_{Sh}$ as the wave function and the equation \eqref{ShrEq} as the Schr\"odinger equation. Note that the equation \eqref{ShrEq} coincides with the equation (11) of Ref. \cite{Zhou:2015}, which  was obtained from the non-relativistic expansion of the the Schr\"odinger equation for two Dirac particles. So, the leading and next-to-leading orders of the non-relativistic expansion for the BSE and Schr\"odinger equation for two interacting Dirac particles coincides. The difference between the expansions of these equations appears in the higher orders corrections in the interaction constant.

Let us consider the correction $\delta\varkappa$ which is  related with the Hamiltonian \eqref{ShrEq} to the ground state energy $\varkappa_{0,0}$ for Coulomb potential, see Eq.~\eqref{KappaFirst}:
\begin{eqnarray}\label{kappaNaivCorrection}
\delta \varkappa_{0,0}=\int d^2r\psi_{0}^+(\bm r)\hat{H}_1\psi_{0}(\bm r),
\end{eqnarray}
where $\psi_{0}(\bm r)$ is the solution of the equation \eqref{SchordFirst} for the case $n=0$ and $l=0$:
\begin{eqnarray}
\psi_0(r)=\frac{\sqrt{2}}{\sqrt{\pi}a_B}e^{-r/a_B}\,,\label{PSI0}
\end{eqnarray}
where $a_B=1/(m\alpha)$ is the Bohr radius. Since the wave function $\psi_0(r)$ equals to constant at $r=0$, the integral \eqref{kappaNaivCorrection} diverges at small distances as:
\begin{eqnarray}
\delta\varkappa_{0,0}\propto\int_0 \frac{dr}{r^2}\,.
\end{eqnarray}
It means that the correction can not be calculated using this expansion. Note that the similar correction which appears in the non-relativistic expansion of the three dimensional Dirac equation is finite \cite{Berestetski}. The reason of the appearance of divergence is that in our case electrons propagate in two spatial dimensions  whereas electric field propagates in three dimension. Therefore, in our case the interaction potential diverges as $1/r$ at $r\to 0$, see \eqref{CoulPot}, whereas the Coulomb potential in two spatial dimensions  diverges only as $\log r$.

The divergence of the integral \eqref{kappaNaivCorrection} indicates that the correction comes from the distances $r \ll a_B$. So, the assumption that the main contribution to the integrals in Eqs. \eqref{fEq}-\eqref{dEq} comes from the momentum scales $p\ll m$ is incorrect for the correction under consideration.  Therefore, to find the correction of the order of $\alpha^4$ we should take into account that the interaction potential differs from the Coulomb potential at the distances $r\sim 1/m$. So, we should consider the distances $p\sim m$ in the system  \eqref{fEq}-\eqref{dEq} more carefully.

\subsection{The correction of the order of $\alpha^4$\label{CorrectExpansion}}
To find the correction of the order of $\alpha^4$ for the $s$-states ($l=0$) we substitute the functions $f(\bm p)$, $h(\bm p)$, and $d(\bm p)$ in the form
\begin{eqnarray}
f(\bm p)=f(p), \,\, h(\bm p)=h(p)e^{2i\phi}, \,\,d(\bm p)=d(p)e^{i\phi}
\end{eqnarray}
to the Eqs.~\eqref{fEq} - \eqref{dEq}, and obtain:
\begin{eqnarray}
(E-2m)f(p)-\frac{p^2}{m}(f(p)+h(p))&=&-\frac{m}{\omega(p)}\int\frac{d\bm k}{(2\pi)^2}V(\bm k-\bm p) f(k)\nonumber\\&-&\frac{p}{2\omega(p)m}\int\frac{d\bm k}{(2\pi)^2}V(\bm k-\bm p)\,k(f(k)+h(k))\cos(\phi_{\bm k}-\phi_{\bm p}),\label{Eqf1}\\
(E+2m)h(p)+\frac{p^2}{m}(f(p)+h(p))&=&\frac{m}{\omega(p)}\int\frac{d\bm k}{(2\pi)^2}V(\bm k-\bm p) h(k)\cos(2(\phi_{\bm k}-\phi_{\bm p}))\nonumber\\&+&\frac{p}{2\omega(p)m}\int\frac{d\bm k}{(2\pi)^2}V(\bm k-\bm p)\,k(f(k)+h(k))\cos(\phi_{\bm k}-\phi_{\bm p})\label{Eqh1}\,,\\
d(p)&=&-\frac{p}{\sqrt{2}m}(f(p)+h(p))\,.\label{Eqd}
\end{eqnarray}
To find the Eqs.~\eqref{Eqf1}-\eqref{Eqd} we express the last two terms in the right-hand side of the  Eq.~\eqref{dEq} using the Eqs.~\eqref{fEq} and \eqref{hEq}, then substitute the result for the function $d(p)$ to the Eqs.~\eqref{fEq} and \eqref{hEq}. Since the Eqs.~\eqref{Eqf1}-\eqref{Eqh1} do not contain the function $d$, below we consider only the functions $f(p)$ and $h(p)$.  The function $d(p)$ can de find using Eq.~\eqref{Eqd}.

One can see that the right-hand sides of the equations \eqref{Eqf1}, \eqref{Eqh1} contain non-hermitian operators $(1/\omega(p))\int d\bm k V(\bm p-\bm k)$. For the further consideration we substitute  the functions $f$ and $h$ in the form:
\begin{eqnarray}
f(p)=\sqrt{\frac{m}{\omega(p)}}a(p),\quad h(p)=\sqrt{\frac{m}{\omega(p)}}b(p),
\end{eqnarray}
then we substitute the energy $E$ in the form \eqref{Ekappa}, and obtain following equations for the functions $a$ and $b$:
\begin{eqnarray}
\varkappa a(p)&=&\frac{p^2}{m}(a(p)+b (p))-\int\frac{d\bm k}{(2\pi)^2}\tilde{V}(\bm p,\bm k) a(k)-\frac{1}{2m^2}\int\frac{d\bm k}{(2\pi)^2}\tilde{V}(\bm p,\bm k)\,(\bm p \bm k)(a(k)+b(k)),\label{Eqa}\\
b(p)&=&-\frac{\varkappa}{4m}b(p)-\frac{p^2}{4m^2}(a(p)+b(p))+\frac{1}{4m}\int\frac{d\bm k}{(2\pi)^2}\tilde{V}(\bm p,\bm k)\frac{(\bm p \bm k)^2-[\bm p\times\bm k]^2}{p^2k^2} b(k)\nonumber\\&+&\frac{1}{8m^3}\int\frac{d\bm k}{(2\pi)^2}\tilde{V}(\bm p,\bm k)(\bm p \bm k)(a(k)+b(k))\label{Eqb}\,,
\end{eqnarray}
where
\begin{eqnarray}
\tilde{V}(\bm p,\bm k)=\frac{m }{\sqrt{\omega(p)\omega(k)}}V(\bm p-\bm k).
\end{eqnarray}
The operator $\tilde{V}$ obeys the condition $\tilde{V}(\bm p,\bm k)=\tilde{V}(\bm k,\bm p)$, therefore the operators in the right-hand side of the Eqs.~\eqref{Eqa}, \eqref{Eqb} are hermitian one.
Also, the potential $\tilde{V}$ decreases faster than potential $V(\bm p)$ at large momentum $p$.
This property allows us to calculate the correction of the order of $\alpha^4$ to the energy $\varkappa_0$.

To calculate the correction we expand the equations \eqref{Eqa} and \eqref{Eqb}. In the necessary order the function
$$b(p)=-\frac{p^2}{4m^2}a(p).$$
We substitute this relation to Eq.~\eqref{Eqa}, retain the terms of the order of $p^4/m^2$, and obtain the equation for the function $a(p)$:
\begin{eqnarray}
\varkappa a(p)=\frac{p^2a(p)}{m}-\frac{p^4a(p)}{4m^3}-\int\frac{d\bm k}{(2\pi)^2}\tilde{V}(\bm p,\bm k) a(k)-\frac{1}{2m^2}\int\frac{d\bm k}{(2\pi)^2}\tilde{V}(\bm p,\bm k)\,(\bm p \bm k)a(k).\label{Eqa1}
\end{eqnarray}
The Eq. \eqref{Eqa1} can be solved using standard methods of the perturbation theory. To find the binding energy in the case of the Coulobm potential \eqref{CoulPot} we present $\varkappa$ and $a$ in the form
\begin{eqnarray}
\varkappa &=&\tilde{\varkappa}+\delta\varkappa,\\
a(p)&=&\tilde{a}(p)+\delta a(p),
\end{eqnarray}
then substitute these representations in Eq.~\eqref{Eqa1}, and obtain:
\begin{eqnarray}
\tilde{\varkappa} \tilde{a}(p)&=&\frac{p^2\tilde{a}(p)}{m}-\int\frac{d\bm k}{(2\pi)^2}\tilde{V}(\bm p,\bm k) \tilde{a}(k),\label{Eqa2}
\end{eqnarray}
\begin{eqnarray}
\delta\varkappa&=&-\frac{\tilde{\varkappa}}{4m^2}\int\frac{d\bm p}{(2\pi)^2}p^2\tilde{a}^2_0(p)-\frac{\tilde{\varkappa}}{4m}\int\frac{d\bm p}{(2\pi)^2}\tilde{a}_0(p)\int\frac{d\bm k}{(2\pi)^2}\tilde{V}(\bm p,\bm k)\tilde{a}_0(k)\nonumber\\
&-&\frac{1}{4m}\int\frac{d\bm p}{(2\pi)^2}\int\frac{d\bm q}{(2\pi)^2}\tilde{V}(\bm p,\bm q)\tilde{a}_0(q)\int\frac{d\bm k}{(2\pi)^2}\tilde{V}(\bm p,\bm k)\tilde{a}_0(k)\nonumber\\
&-&\frac{1}{2m^2}\int\frac{d\bm p}{(2\pi)^2}\tilde{a}_0(p)\int\frac{d\bm k}{(2\pi)^2}\tilde{V}(\bm p,\bm k)(\bm p\bm k)\tilde{a}_0(k),\label{EqTildeKappa}
\end{eqnarray}
where the function
$$\tilde{a}_0(p)=\int d\bm r e^{-i (\bm r\bm p)}\psi_0(r)=\frac{2\sqrt{2\pi}a_B}{(1+p^2 a_B^2)^{3/2}}.$$
The function $\psi_0(r)$ is the solution of the Schr\"odinger equation for the particle in the Coulomb potential \eqref{PSI0}.

%
To obtain Eq.~\eqref{EqTildeKappa} we have used the Eq.~\eqref{Eqa2}. The analytical calculation of the correction $\delta \varkappa$ and quantity $\tilde{\varkappa}$ up to the terms of the order of $\alpha^4$  gives:
\begin{eqnarray}
\tilde{\varkappa}&=&-m\alpha^2+m\alpha^4\left(-{4}+\frac{\pi}{2}+3\log 2+2\log\frac{1}{\alpha}\right),\\
\delta\varkappa&=&-\frac{m\alpha^4}{4}\left(1+\pi+6\log2+4\log\frac{1}{\alpha}\right).
\end{eqnarray}
So, the ground state energy in the leading and next-to-leading order in the parameter $\alpha$ has the form:
\begin{eqnarray}\label{kappaAlpha4}
\varkappa=\tilde{\varkappa}+\delta\varkappa=-m\alpha^2
\left(1-\alpha^2\left[\log\frac{1}{\alpha}-\frac{17}{4}+\frac{\pi}{4}+\frac{3}{2}\log 2\right]\right).
\end{eqnarray}
The first term in brackets coincides with the energy \eqref{KappaFirst}, the second term is the correction. One can see that the $\alpha^4$ corrections contain the logarithmic term $\log\alpha$ which comes from the momentum region $a_B^{-1}\ll p\ll m$. It means that it is necessary to take into account the corrections to the interaction potential related with electron-hole exitations in the intermediate states. 
The calculation of the correction for the first excited state ($2s$-state) gives
\begin{eqnarray}\label{kappaAlpha4_2s}
\varkappa_{2s}=-\frac{m\alpha^2}{9}\left(1-\frac{\alpha^2}{3}\left[\log\frac{12}{\alpha}+\frac{\pi}{2}-\frac{23}{4}\right]\right)\,.
\end{eqnarray}
One can see, that it also contain the logarithmic term $\log\alpha$. The corrections of the order of $\alpha^4$ to the other states can be calculated using described procedure.  

\begin{figure}[t]
\includegraphics[width=7cm]{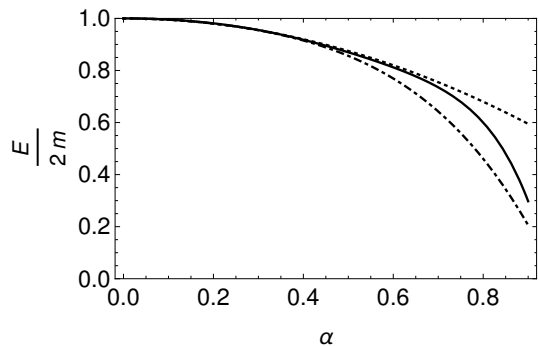}
\caption{\label{Fig1} The dependence of the ratio $\frac{E}{2m}$ on $\alpha$ for the ground state. The solid line, the dotted line, and the dashed-dotted line correspond to the exact solution of the BSE, the solution of the Schr\"odinger equation \eqref{KappaFirst}, and the result  Eq.~\eqref{kappaAlpha4}, respectively.}
\end{figure}

So, the BSE \eqref{Eqf1}-\eqref{Eqd} allows us to find the correction of the order of $\alpha^4$, whereas the equation \eqref{ShrEq} does not give the possibility to calculate the correction, since it does not take into account the difference of the potential form the Coulomb potential at the distances $r\sim 1/m$. The results \eqref{kappaAlpha4}  and \eqref{kappaAlpha4_2s} are shown in Figs.~\ref{Fig1} and \ref{Fig3} by the dashed-dotted line. One can see that the corrections are in a good agreement with the exact numetical solution of the BSE for the parameters $\alpha\lesssim 0.5$. For the case $\alpha>0.5$ it is necessary to take into account the $\alpha^6$ terms.
\begin{figure}[t]
\includegraphics[width=7cm]{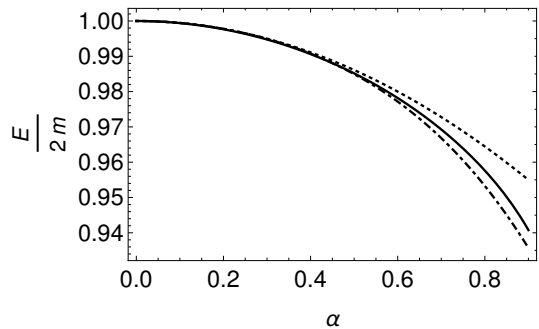}
\caption{\label{Fig3} The dependence of the ratio $\frac{E}{2m}$ on $\alpha$ for the $2s$ state. The solid line, the dotted line, and the dashed-dotted line correspond to the exact solution of the BSE, the solution of the Schr\"odinger equation \eqref{KappaFirst}, and the result  Eq.~\eqref{kappaAlpha4_2s}, respectively.}
\end{figure}

Strictly speaking, there are another correction of the order of $\alpha^4$ which we do not take into account. These corrections are related with the Uehling potential,  the cross-box diagram contribution and annihilation diagram contribution. Here we do not consider these corrections and concentrate our attention only on the BSE in the leading order in the interaction potential. The contribution of the mentioned corrections will be consider in the following papers.

\section{Exact solutions of the Bethe-Salpeter equation for the case of $l=0$\label{BetheSalpeterNumeric}}
In the previous section we obtain that the correction of the order of $\alpha^4$ is not small therefore we solve the BSE exactly for the case of the Coulomb potential and for the states with zero angular momentum $l=0$. For this case it is convenient to present the function $\psi(\bm p)$ in the following form:
\begin{eqnarray}\label{psiL=0}
\psi(E,\bm p)=\frac{F(p)+H(p)}{\sqrt{2}}|1,1\rangle + e^{2i\phi_{\bm p}}\frac{F(p)-H(p)}{\sqrt{2}}|1,-1\rangle + e^{i\phi_{\bm p}}d(p)|0,0\rangle\,.
\end{eqnarray}
The substitution of the function $\psi(\bm r)$ in the form \eqref{psiL=0} to the Eq. \eqref{BetheSalpEq} gives the following system of the equations
\begin{eqnarray}
\frac{E}{2m}F(p)&=&H-\frac{1}{2\omega(p)}\int\frac{d\bm k}{(2\pi)^2}V(\bm k-\bm p)\left[F(k)\sin^2\frac{\phi_{\bm k}-\phi_{\bm p}}{2}+H(k)\cos^2\frac{\phi_{\bm k}-\phi_{\bm p}}{2}\right],\label{EqF}\\
\frac{E}{2m}H(p)&=&F+\frac{p^2}{m^2}F-\frac{1}{2\omega(p)}\int\frac{d\bm k}{(2\pi)^2}V(\bm k-\bm p)\left[F(k)\cos^2\frac{\phi_{\bm k}-\phi_{\bm p}}{2}+H(k)\sin^2\frac{\phi_{\bm k}-\phi_{\bm p}}{2}\right]\nonumber\\
&-&\frac{p}{2\omega(p)m^2}\int\frac{d\bm k}{(2\pi)^2}V(\bm k-\bm p)\,kF(k)\cos(\phi_{\bm k}-\phi_{\bm p}),\label{EqH}\\
d(p)&=&-\frac{p}{m}F(p).\label{EqD}
\end{eqnarray}
One can see that the Eq.~\eqref{EqD} is trivial, therefore we can consider only Eqs.~\eqref{EqF} and \eqref{EqH}. Below we solve these equations numerically for two potentials: the Coulomb potential Eq.~\eqref{CoulPot}, and the localized potential without singularity at the point $r=0$:
\begin{eqnarray}
U(r)=U_0 e^{-r^2/R^2},\label{ExpPotential}
\end{eqnarray}
where $U_0$ is the depth of the potential well, $R$ is the width of the potential well. In the momentum space the potential has the form:
\begin{eqnarray}
U(\bm p)=\pi U_0R^2\exp\{-R^2 p^2/4\}.
\end{eqnarray}

\subsection{Coulomb potential}
\begin{figure}[b]
\includegraphics[width=7cm]{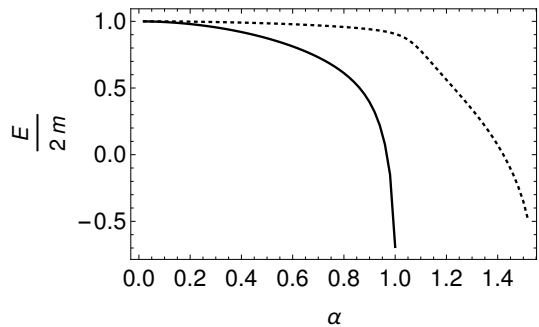}
\caption{\label{Fig2} The dependence of the ratio ${E}/{2m}$ on $\alpha$. The solid line corresponds to the ground state.  The dotted line corresponds to the  first excited state with $n=1$, $l=0$.}
\end{figure}

We start our consideration with the Coulomb potential. The Coulomb potential decreases only as $1/q$ at large $q$, therefore to increase the accuracy of the numerical calculation we extract the asymptotic behavior of the functions  $F(p)$ and $H(p)$ analytically at large momentum $q\gg m$. The substitution of the functions $F(p)$ and $H(p)$ in the form
\begin{eqnarray}
F(p)=a/p^{2\beta},\quad H(p)=b/p^{2\beta_1}
\end{eqnarray}
gives the following equations for  the $\beta$ and $\beta_1$ in the leading order in the parameter $\alpha$:
\begin{eqnarray}
\frac{\Gamma\left(\beta-1/2\right)\Gamma(2-\beta)}{\Gamma(\beta)\Gamma(5/2-\beta)}=\frac{4}{\alpha},\\
\frac{\Gamma\left(\beta_1-1/2\right)\Gamma(1-\beta_1)}{\Gamma(\beta_1)\Gamma(3/2-\beta_1)}=\frac{8}{\alpha}.
\end{eqnarray}
In the leading order in the parameter $\alpha$ we obtain the solutions of the equations:
\begin{eqnarray}
\beta\approx 2-\frac{\alpha}{8},\quad\beta_1\approx 1-\frac{\alpha}{8}.
\end{eqnarray}
Therefore, for the numerical calculations we substitute the functions $F(p)$ and $H(p)$  to the Eqs.~\eqref{EqF} and \eqref{EqH} in the form:
\begin{eqnarray}
F(p)=\frac{ \tilde{F}(p)}{(m^2+p^2)^{\beta}}\,, \quad H(p)=\frac{\tilde{H}(p)}{(m^2+p^2)^{\beta_1}}\,,
\end{eqnarray}
where $\tilde{F}(p)$ and $\tilde{H}(p)$ tends to constant at $p\gg m$. The solution $E/2m$ for different values  of the parameer $\alpha$ is presented in Figs. \ref{Fig1} and \ref{Fig2}. In Fig.~\ref{Fig1} the dependence of the ground state energy in the units $2m$ on the interaction constant is plotted. One can see that the exact solution of the BSE significantly differs from the result \eqref{KappaFirst} only for $\alpha>0.7$.  However, for $\alpha>0.8$ the difference is sufficient. Moreover, the critical value of $\alpha_c$  ($E(\alpha_c)=-2m$) for  BSE differs from the critical value of the interaction constant for the Sch\"odinger equation. The  ground state energy $E$ for BSE  achieves the valence band ($E=-2m$) at $\alpha=1\pm0.05$. So, with our calculation precision, we obtain that the critical value of the interacting constant $\alpha_c=1$. The uncertainty is related with the dependence of the result on the sampling step when calculating the integrals. The results  $E/(2m)$ for $\alpha<0.9$ is weakly depend on the discretization step of the numerical integration. For $\alpha>\alpha_c$ the energy level of the ground state disappears from the discrete spectrum, see Fig.~\ref{Fig2}. 
In Fig.~\ref{Fig2} we plot the dependence $E(\alpha)$ for the ground state and the first excited state with $n=1$, $l=0$. One can see that  when the $\alpha$ is approaching to unity the slope of the function $E(\alpha)$  goes to $-\infty$. For $\alpha>1$  the state with $n=1$, $l=0$ becomes the ground state. Such behavior is similar to that of electron in the Coulomb field in the three dimensional electrodynamics.

\subsection{Exponential potential}
To investigate the behavior of the energy levels in the vicinity of critical value of the interacting constant in detail we consider the toy potential \eqref{ExpPotential}. In Fig.~\ref{Fig3} the dependence $E/2m$ on $U_0/m$ for three lowest binding energies of BSE is presented.
\begin{figure}[t]
\includegraphics[width=7cm]{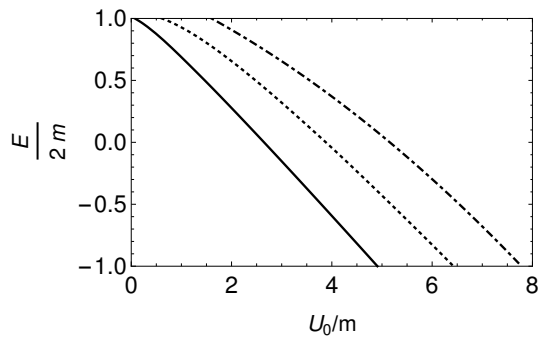}
\caption{\label{Fig3} The dependence of the ratio $\frac{E}{2m}$ on $U_0$. The solid line corresponds to the ground state.  The dotted line corresponds to the  first excited state with $n=1$, $l=0$, and dash-dotted line corresponds to $n=2$, $l=0$ state.}
\end{figure}
One can see that when the interaction constant is increasing the energy level  is decreasing then it cross the line $E=0$ and then riches the valence band $E=-2m$ at $\tilde{U}=\tilde{U}_c$. For the exponential potential \eqref{ExpPotential} the slope of the function $E(\tilde{U})$ does not go to $-\infty$ when $\tilde{U}$ in the vicinity of $\tilde{U}_c$, because of the potential does not have singularities at $r\to 0$. Such behavior resembles the behavior of the electron in the potential well in three dimensional quantum electrodynamics.

\section{Conclusion}
In the present paper we have investigated the electron-hole interaction in the transition-metal dichalcogenides using the BSE obtained in the leading order in the potential. We have obtained the equation for the four component  wave function of the exciton. We have investigated the dependence of a few lowest binding energies on the interaction constant. For small interaction constant the equation \eqref{Eqa1} and  analytical expression for the ground state energy have been found  up to the terms of the order of  $\alpha^4$.   We have demonstrated that the expansion of BSE in the leading order in the interaction constant coincides with expansion of the Schr\"odinger equation for the two interacting Dirac particles. However, in the next-to-leading order in the interaction constant the equations are different. The expansion of the Schr\"odinger equation leads to the appearance of the operators with non-integrable singularities. The appearance of the singularities means that even for small $\alpha$ at small distances the problem is ''relativistic''. It indicates that the distances $r\ll a_B\sim m/\alpha$ give the contribution to correction for the energy levels of exciton. The correct expansion of the BSE does not contain the singular operators. Using the equation we have calculated the correction to the binding energies and demonstrated that it contains the term $\alpha^2\log\alpha$ which indeed comes from the distances $(1/m)\ll r\ll a_B$.

We have investigated the binding energies for the Coulomb potential and for the exponential potential \eqref{ExpPotential}. We have obtained that for the Coulomb potential the value $\alpha=1$ is critical. When the parameter $\alpha$ approaches to unity the energy of the ground state goes to zero. If the value of the parameter $\alpha$ is slightly greater than $1$ this state disappears from the discrete spectrum. This picture is close to that in quantum electrodynamics, see \cite{Berestetski}. The reason of such behavior is the singularity of the Coulomb potential. To confirm this statement we found the spectrum for the potential without singularities. We showed that for the increasing interaction constant the bound state energies decreases from $E=2m$ smoothly to $E=0$, and then to $E=-2m$. The energy level disappears when it reaches the valence band ($E=-2m$).

We have obtained that when the interaction constant $\alpha$ is close to $0.5$-$0.6$ the binding energies slightly differ from that obtained in the leading order, see Eq.~\eqref{KappaFirst}. So, the BSE obtained in the leading order does not describe the experimental results. However, the investigation of the BSE shows that the structure of the potential at small distances should  affect energy levels significantly. Therefore, to describe the experimental data it is necessary to include the Uehling type diagrams, and cross-box type of the diagrams  in the BSE. These types of the corrections will be considered in the following papers.
%

\acknowledgments
I.S.T. and P.A.K. grateful to prof. O. P. Sushkov for drawing attention to the problem. 


\begin{thebibliography}{99}
\bibitem{Novoselov:2004}
K. S. Novoselov et al., Science 306, 666 (2004).

\bibitem{Zhu:2011}
Z. Y. Zhu, Y. C. Cheng, and U. Schwingenschl\"ogl, Phys. Rev. B \textbf{84}, 153402 (2011).

\bibitem{Xiao:2012}
D. Xiao, G.-B. Liu, W. Feng, X. Xu, and W. Yao, Phys. Rev. Lett. {\bf 108}, 196802 (2012).

\bibitem{Rodin:2013}
S. A. Rodin, A. H. Castro Neto, Phys. Rev. B \textbf{88}, 195437 (2013).


\bibitem{Zhou:2015}
J. Zhou, W.-Y. Shan, W. Yao, and D. Xiao, Phys. Rev. Lett. {\bf 115}, 166803 (2015).

\bibitem{Trushin:2016}
M. Trushin, M. O. Goerbig, and W. Belzig, Phys. Rev. B {\bf 94}, 041301 (2016).

\bibitem{Trushin:2018}
M. Trushin, M. O. Goerbig, and W. Belzig, Phys. Rev. Lett. {\bf 120}, 187401 (2018).

\bibitem{Trushin:2019}
M. Trushin, Phys. Rev. B {\bf 99}, 205307 (2019).

\bibitem{Glinskii:1987}
 G. F. Glinskii and Zl. Koinov,  Theor. Math. Phys. {\bf 70}, 252 (1987).

\bibitem{Scharf:2019}
B. Scharf, D. V. Tuan, I. Zutic, and H. Dery,  J. Phys.: Condens. Matter {\bf 31}, 203001 (2019).


\bibitem{Chernikov:2014}
A. Chernikov, {\it et al.}, Phys. Rev. Lett.   {\bf 113}, 076802 (2014).


\bibitem{Kumar:2014}
A. Chernikov, {\it et al.}, Phys. Rev. Lett.   {\bf 113}, 026803 (2014).


\bibitem{Berestetski}
V. Berestetski, E. Lifshits, and L. Pitayevsky, {\it Quantum electrodynamics} (Pergamon, 1982).

\bibitem{Itzykson}
C. Itzykson, and J.-B. Zuber, {\it Quantum field theory} (McGraw-Hill, 1980).







\end{thebibliography}
\end{document}